\documentclass[prl,twocolumn,floatfix]{revtex4}
\usepackage{array,amsmath,amssymb,graphicx}

\begin{document}

\title{Ground-state energy of the electron liquid in ultrathin wires}

\author{Michael M. Fogler}

\affiliation{Department of Physics, University of
California San Diego, La Jolla, California 92093}

\date{\today}

\begin{abstract}

The ground-state energy and the density correlation function of the
electron liquid in a thin one-dimensional wire are computed. The
calculation is based on an approximate mapping of the problem with a
realistic Coulomb interaction law onto exactly solvable models of
mathematical physics. This approach becomes asymptotically exact in the
limit of small wire radius but remains numerically accurate even for
modestly thin wires.

\end{abstract}

\maketitle

Recently much attention has been devoted a class of one-dimensional (1D)
conductors that can be termed ultrathin wires. Examples of such systems
include single-wall carbon nanotubes (CN)~\cite{Saito_book},
semiconductor nanowires~\cite{Huang_01}, and conducting
molecules~\cite{Heath_03}. {\it Semiconducting\/} ultrathin wires are
especially interesting because their electron concentration $n$ can be
varied by the field-effect, which can be used for creating miniature
electronic devices~\cite{Tans_98, Javey_02}. In such applications a
fundamental role is played by the concentration-dependence of the
ground-state energy density $\varepsilon(n)$. This function determines
the electrostatic screening and affects the capacitive coupling of the
electron liquid to external voltage sources. It is also a core input of
the density-functional theory (DFT), which is the basis of today's
electronic structure calculations. Since the 3D Fermi-liquid theory does
not hold in 1D, it is unclear whether the usual DFT optimized for
three-dimensional (3D) systems is adequate for ultrathin wires. The
Luttinger liquid (LL) theory~\cite{Haldane_81}, which is called upon to
replace the Fermi-liquid theory, makes no predictions for the
short-range physics that determines $\varepsilon(n)$. Therefore, the
calculation of the ground-state energy of 1D wires with realistic
Coulomb interactions has remained an open problem. The primary
difficulty is the computation of the correlation energy
$\varepsilon_{cor}$, which is determined by the shape and size of the
exchange-correlation hole (XCH), i.e., the reduction in probability of
any two electrons closely approaching each other. Below we propose a
theory that calculates these quantities.

\noindent{\it Model\/}.--- Our calculation is done for an $N$-component
electron gas, $N$ being the combined spin-valley degeneracy of the
electron spectrum. For example, $N = 4$ in CN~\cite{Saito_book}. The
aforementioned XCH is the term that refers to the negative dip of the
two-body cluster function $h(x)$ around $x = 0$. Here $h(x) = (M n)^{-1}
\sum_{i \neq j} \langle \delta(x_i - x_j - x) \rangle - 1$, $M$ is the
number of electrons, and $x_i$ are their coordinates. Larger $|h(0)|$
imply stronger correlations. Since $\int_{-\infty}^{\infty} h(x) dx = -1
/ n$, the XCH has a characteristic width $l_* \sim 1 / n |h(0)|$. For
example, in the free Fermi gas $h(x) = -N \sin^2 (n x / N) / n^2 x^2$,
so that $|h(0)| = 1 / N$ and $l_*$ is equal to $N / n$, the average
distance among electrons of same species or, as we call it, same {\it
isospin\/}. Our goal is to compute $h(x)$ for an interacting system.
Once $h(x)$ is known, $\varepsilon(n)$ can be obtained
straightforwardly, see below. We model the interactions by the potential
$U(x) = e^2 / \kappa \sqrt{x^2 + R^2}$, which accounts for smoothing of
Coulomb repulsion at distances of the order of the wire radius $R$. The
wire is considered ultrathin if the parameter ${\cal L} = \ln(a_B / R)$
is large, where $a_B = \hbar^2 \kappa / m e^2$, $m$, and $\kappa$ are
the effective Bohr radius, electron mass, and dielectric constant,
respectively. On general grounds, we may expect that at low densities,
$n \ll 1 / a_B$, electrons should form a 1D Wigner
``crystal''~\cite{Schulz_93} with $h(x)$ sharply peaked at integer
multiples of $a = 1 / n$. At $n \gg 1 / a_B$ where electrons have a
large kinetic energy, $h(x)$ should remain appreciable down to $x \ll
a$. Below we refine and flesh out this qualitative picture by
quantitative calculations.

Crucial for our approach is the fact that to the leading order in $1 /
{\cal L}$ the problem in hand and the problem with the contact
interaction, $U(x) = (\hbar^2 c / m) \delta(x)$ give the same
short-range behavior of the correlation functions, including the
XCH. Here $c$ is given by
\begin{equation}
                  c = ({2} / {a_B}) \ln ({l_*} / {R}).
\label{c}
\end{equation}
This remarkable mapping between the two interaction laws holds only in
the liquid state, $n \gg 1 / a_B$. The reason for it becomes clear if
one carefully separates the effects of the sharp maximum (``core'') of
the Coulomb potential $U(x)$ at $x = 0$ from those of its $1 / x$ tails.
As was shown in our earlier work~\cite{Fogler_04}, the condition $n \gg
1 / a_B$ guarantees that the Coulomb tails have negligible effects on
$h(x)$ up to exponentially large distances, $\ln(x / a) \sim 1 / r_s$,
where $r_s = a / 2 a_B \ll 1$. Since $r_s$ plays the role of the
dimensionless coupling constant, this agrees with the conventional
wisdom. On the other hand, the electron scattering caused by the
short-range core of $U(x)$ is enhanced~\cite{Fogler_04} by the large
logarithm ${\cal L}$. Therefore, the Coulomb potential acts as a sum of
a strong short-range core and weak tails, and so can be mapped onto a
suitable $\delta$-function.

The rest of the paper is organized as follows. We begin by studying
certain limiting cases, which verify the correctness of our
choice~(\ref{c}) of the coefficient $c$. We then explain how our theory
can be used to calculate $\varepsilon(n)$ at all $r_s \ll 1$. We proceed
to the study of the large-$r_s$ Wigner crystal where the mapping onto
the contact interaction model is no longer valid. We show that the exact
asymptotics of $\varepsilon(n)$ in the $r_s \gg 1$ limit can
nevertheless be derived while at $r_s \sim 1$ a simple variational
approximation can be used. We also present a numerical scheme that
unifies all the asympotical formulas we obtain. It yields a seemless
interpolation over the entire range of $n$ even for ${\cal L} \sim 2-3$.
We take it as evidence that our theory remains numerical accurate even
for modestly thin wires, which may stimulate its use in practical DFT
calculations. (Achieving large ${\cal L}$ is feasible~\cite{Fogler_04}
but technically difficult).

\noindent{\it Definitions\/}.--- We do the usual subtration of the
Hartree term in the definition of the energy density, $\varepsilon(n)
\equiv L^{-1} [\langle H \rangle - \tilde{U}(0) n^2 / 2]$, where $L$ is
the length of the wire, $H$ is the Hamiltonian, and tilde denotes the
Fourier transform. We further define the correlation energy density
$\varepsilon_{cor}$ as the difference between $\varepsilon$ and the sum
of the kinetic $\varepsilon_0$ and the exchange $\varepsilon_{x}$
energies of the Fermi gas,
\begin{equation}
\varepsilon_0 = \frac{\pi^2}{6} \frac{\hbar^2}{m} \frac{n^3}{N^2},
\quad \varepsilon_{x} \simeq -\frac{e^2}{\kappa} \frac{n^2}{N}
 \left(\ln \frac{N}{R n} + A_T\right), 
\label{Fermi_gas}
\end{equation}
where $A_T = \frac32 - \gamma - \ln \pi \approx -0.222$, $\gamma$ is
Euler constant\cite{Gradshteyn_Ryzhik}. 

The relations among $\varepsilon_{cor}(n)$, $h(x)$, and the
dielectric function $\epsilon(q,\omega)$ are (see, e.g.,
Ref.~\onlinecite{Mahan_book}, Secs.~5.4 and 5.6):
\begin{eqnarray}
&\displaystyle  \varepsilon_{xc} \equiv \varepsilon_{x}
 + \varepsilon_{cor} = n^3 \int\limits_0^{r_s} \frac{d r_s}{r_s}
\frac{\varepsilon_{int}(n, r_s, N)}{n^3},&
\label{xc_energy}\\
&\displaystyle  \varepsilon_{int} = n^2 \int\limits_0^\infty d x U(x) h(x),&
\label{interaction_energy}\\
&\displaystyle \tilde{h}(q) = -1 - \frac{\hbar}{n \tilde{U}(q)}
 \int\limits_0^\infty
 \frac{d \omega}{\pi} \,\text{Im} \left[\frac{1}{\epsilon(q, \omega)}
 \right].&
\label{h_from_epsilon}
\end{eqnarray}

\noindent{\it RPA regime\/}.--- The validity of our mapping between the
Coulomb and the contact interactions can be verified by an independent
method if the limit of large $N$ (actually, large $N^2$) is taken. We
discuss it because it is not only an instructive example but also the
case relevant for CN, where $N^2 = 16$. For large $N$, $\epsilon(q,
\omega)$ is dominated by the random-phase approximation
(RPA)~\cite{Mahan_book}, which sums order-by-order the diagrams with the
largest number of fermion loops. For $q \gg k_F = \pi n / N$ the result
is
\begin{equation}
\epsilon(q, \omega) = 1
 + \frac{2 n E(q) \tilde{U}(q)}{E^2(q) - (\hbar \omega + i 0)^2},
\quad E(q) = \frac{\hbar^2 q^2}{2 m}.
\label{epsilon_RPA}
\end{equation}
Combined with Eq.~(\ref{h_from_epsilon}), it entails that at ${\cal L}
\ll n a_B \ll N^2 {\cal L}$ (the RPA regime), the XCH has the depth
$|h(0)| \simeq (\pi n l_*)^{-1}$ and a characteristic width
\begin{equation}
          l_* = \sqrt{a_B / 2 n \ln (l_* / R)}.
\label{l}
\end{equation}
The XCH is much deeper than in the Fermi gas, $|h(0)| \gg 1 / N$, and so
the correlations are strong; yet $|h(0)| \ll 1$, so that the RPA is
still reliable. From Eq.~(\ref{interaction_energy}) we find, to the
leading order in $1 / N$,
\begin{equation}
  \varepsilon_{xc} \simeq -\frac{2}{3\pi} \frac{e^2}{\kappa a_B^2}
\left\{\frac{1}{r_s}\left[
\ln \left(\frac{l_*}{R}\right) - \frac12\right]\right\}^{3/2}.
\label{xc_RPA}
\end{equation}
%
%
Repeating the same calculation for the contact interaction with $c$
given by Eq.~(\ref{c}), we obtain {\it exactly the same\/} result. To
track down how this comes about it is convenient to do the integration
in Eq.~(\ref{interaction_energy}) in the $q$-space. The interaction
potential enters through its Fourier transform $\tilde{U}(q) \simeq (2
e^2 / \kappa) \ln (1 / q R)$, which is a slow function of $q$. The
integral is dominated by $q \sim 1 / l_*$, and so to the leading order
in ${\cal L}^{-1}$, $\tilde{U}(q)$ can be replaced by the $\tilde{U}(1 /
l_*)$, i.e., $U(x) \to (\hbar^2 c / m) \delta(x)$, as we claimed above.

The RPA eventually breaks down at small $n$, where it predicts $h(0)$ to
drop below the strict lower bound of $-1$ required by the nonnegativity
of the electron density. This places the lower boundary of the RPA
regime at $n \sim {\cal L} / \pi a_B$. What happens at lower $n$ is
discussed next.

\noindent{\it CTG regime.\/}--- The case of $n \ll {\cal L} / a_B$ has
in fact already been studied in Ref.~\cite{Fogler_04}. We showed that at
such $n$ electrons should form a correlated state of the Coulomb Tonks
Gas (CTG). The CTG can be defined as the state where on short
lengthscales electrons behave as impenetrable but otherwise free. It
owes its name to a certain similararity it enjoys with the
Tonks-Girardeau gas of 1D cold atoms~\cite{Paredes_04}. It is worth
mentioning that the long-distance behavior in the RPA, CTG, and Wigner
crystal regimes is universally the same and is described by the LL
theory. In the limit $R \to +0$, i.e., $c, {\cal L} \to \infty$, the
ground-state wavefunction $\Psi$ factorizes into the isospin part $\Phi$
and the orbital part (the remainder)~\cite{Fogler_04}:
\begin{equation}
\Psi = \Phi \times e^W (-1)^Q \prod\limits_{Qi > Qj}
        \left[\sin \frac{\pi}{L}(x_{Qi} - x_{Qj})\right]^\lambda,
\label{Psi}
\end{equation}
where $Q1$ through $Q{M}$ are the indices in the spatially ordered list
of the electron coordinates $0 < x_{Q1} < \ldots < x_{Q{M}} < L$
(periodic boundary conditions are assumed), $(-1)^Q$ is the parity of
the corresponding permutation, and $\lambda = 1$ for now. For $N = 2$,
$\Phi$ coincides with the ground-state of a spin-$1/2$ Heisenberg
chain; for $N > 2$, see Ref.~\onlinecite{Sutherland_75}. We will not
discuss the function $W$ here because it has negligible effect on $h(x)$
for $x \ll a \exp(1 / r_s)$~\cite{Fogler_04}. Once $W$ is set to zero,
$\Psi$ becomes the ground state of the contact-interaction problem at $c
= \infty$ (the gas of impenetrable fermions)~\cite{Ogata_90,
Schlottmann_97}. This is another explicit demonstration of our mapping,
this time in the ${\cal L} \to \infty$ limit. Note that the XCH has the
largest possible depth of unity and the width $l_* = a$.

For a finite $R$, $\Psi$ remains the correct approximation to the ground
state to the leading order in ${\cal L}^{-1}, r_s \ll 1$. We use $\Psi$
(with $W = 0$) as a trial state to evaluate $\varepsilon(n)$.
Independent of the form of $\Phi$, the result is given by
Eq.~(\ref{Fermi_gas}) with $N = 1$ (see also Ref.~\cite{Fogler_04}), 
\begin{equation}
\varepsilon(n) = \frac{e^2}{\kappa} n^2
[\ln (R n) - A_T] + \frac{\pi^2}{6} \frac{\hbar^2}{m} n^3,
\label{energy_Tonks}
\end{equation}
which agrees with $\varepsilon(n)$ for the contact-interaction problem
to the order $1 / {\cal L}$ [Eq.~(\ref{Tonks_energy_Bethe})], validating
our mapping once again.

\noindent{\it Bethe ansatz.\/}--- The most remarkable consequence of the
mapping between the Coulomb and the contact-interaction models is that a
unified treatment of all $r_s \ll 1$ regimes is possible. This is due to
the fact fact that the latter model is solvable by the Bethe
ansatz~\cite{Yang_67}. The exact energy density at any given $n$ is
given by~\cite{Schlottmann_97}
\begin{equation}
  \varepsilon(n) = -\frac{\hbar^2}{2 m} c n^2 + \frac{\hbar^2}{2 m}
                \int\limits_{-Q}^{Q} d k k^2 \rho(k),
\label{energy_Bethe}
\end{equation}
where $\rho(k)$ is the solution of the integral equation
\begin{eqnarray}
  \frac{1}{2\pi} &=& \rho(k) - \int\limits_{-Q}^{Q} d k^\prime
  G(k - k^\prime) \rho(k^\prime),
\label{rho_equation}\\
G(k) &=& \frac{1}{\pi c N} \,\text{Re}\left[
\psi\left(1 + \frac{i k}{N c}\right) -
\psi\left(\frac{i k + c}{N c}\right)\right].\quad
\label{G}
\end{eqnarray}
Here $\psi(z)$ is the digamma function~\cite{Gradshteyn_Ryzhik}
and $Q = Q(n)$ is fixed by the constraint $n = \int_{-Q}^{Q}
d k \rho(k)$. Two analytical asymptotics of the solution
can be obtained~\cite{Schlottmann_97, Fogler_unpublished}
\begin{eqnarray}
\varepsilon &\simeq&  \frac{\hbar^2}{m}\left[
-\frac{2}{3 \pi} (c n)^{3/2} + \frac{\pi^2}{6} \frac{n^3}{N^2}
\right],
\quad c \ll n \ll c N^2,\quad
\label{RPA_energy_Bethe}\\
&\simeq& \frac{\hbar^2}{m}\left[
-\frac{1}{2} c n^2 + \frac{\pi^2}{6} n^3
\right],
\quad n \ll c,
\label{Tonks_energy_Bethe}
\end{eqnarray}
in agreement with Eqs.~(\ref{xc_RPA}) and (\ref{energy_Tonks}). From the
theory point of view, Eqs.~(\ref{c}), (\ref{l}), (\ref{energy_Tonks}),
and (\ref{energy_Bethe})--(\ref{G}) solve the problem of computing
$\varepsilon(n)$ at all $r_s \ll 1$. A {\it practical\/} algorithm for
finding the solution is given shortly below.

\noindent{\it Wigner crystal.\/}--- At very low densities, $r_s \gg 1$,
the mapping onto the contact interaction problem is however {\it
invalid\/}. The tails of the Coulomb barriers that separate nearby
electrons are strong enough to keep them at almost equidistant positions
(although the long-range order is eventually destroyed by fluctuations).
According to the standard strong-coupling perturbation theory, the
ground-state energy in this regime is equal to the Madelung sum plus the
zero-point phonon energy,
\begin{equation}
       \varepsilon = \frac{e^2}{\kappa} n^2 [\ln (R n) - A_W]
                   + C_{ph} \frac{e^2}{\kappa} n^{5/2} a_B^{1/2},
\label{energy_Wigner}
\end{equation}
where $A_W = \ln 2 - \gamma \approx 0.116$ and $C_{ph} \approx 1.018$.
As for the cluster function $h(x)$, it can be obtained by interpolating
between the collective phonon-like correlations at $x \gtrsim a$ and
two-body correlations at $x \ll a$ (see, e.g.,
Ref.~\onlinecite{Zhu_96}).

%
%
\begin{figure}
\centerline{
\includegraphics[width=3.1in,bb=80 145 510 566]{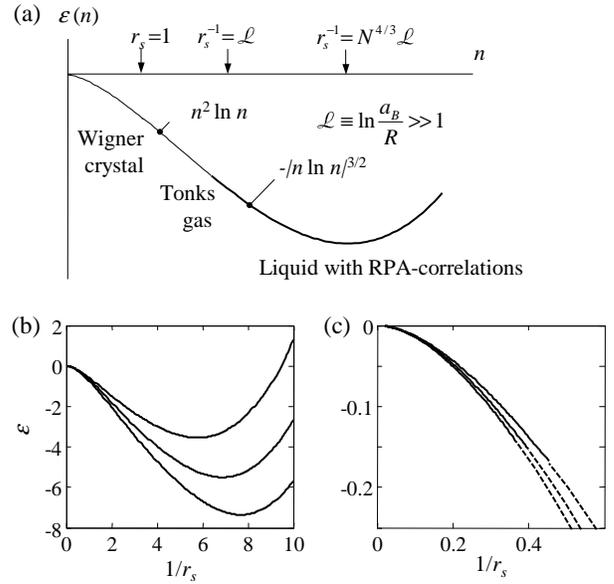}
}
\setlength{\columnwidth}{3.2in}
\caption{
(a) Qualitative behavior of $\varepsilon(n)$. (b)
$\varepsilon$ in units of $e^2 / \kappa a_B^2$ for $a_B / R = 10, 15,
20$ (top to bottom), evaluated numerically. (c) The
low-density part of the same plot; solid lines are from the variational
method, the dashed lines --- from the Bethe ansatz.
\label{Fig_energy}
}
\end{figure}

\noindent{\it Variational and numerical interpolation\/}--- Until now we
expanded on a formalism that gives results both for $\varepsilon(n)$ and
for $h(x)$ that are rigorously correct to the leading order in a
suitable small parameter, either $1 / {\cal L}$ or $1 / r_s$. These
results, e.g., the functional form of $\varepsilon(n)$ in various
regimes [Fig.~\ref{Fig_energy}(a)] have an academic or methodological
interest. In the remainder of the paper we shift the focus to a more
pragmatic goal. We wish to find a computational scheme that gives an
accurate {\it numerical\/} approximation to the same quantities when
neigher ${\cal L}$ nor $r_s$ are truly large. We achieve this by
combining a variational method with a numerical interpolation. Some
results are shown in Fig.~\ref{Fig_energy}(b) and (c). The concrete
interpolation scheme used in generating these plots is as follows. For
$r_s \lesssim 2$, $\varepsilon(n)$ is calculated by numerically solving
Eqs.~(\ref{c}), (\ref{energy_Bethe})--(\ref{G}) with $l_* = \pi \exp(A_T)
/ Q(n, c)$. Note that function $Q(n)$ has the following limiting forms:
$Q \simeq \pi n$ for $n \ll c$, $Q \simeq 2 \sqrt{n c}$ for $c \ll n \ll
c N^2$, and $Q = \pi n / N$ for $n \gg c N^2$. This entails that our
choice of $l_*$ is exact at small and large $n$, and is adequate
everywhere in between, see Eqs.~(\ref{Fermi_gas}), (\ref{l}), and
(\ref{energy_Tonks}).

All what remains is to handle $r_s \gtrsim 2$ regime where the crossover
between the CTG and the Wigner crystal occurs. Our solution is to treat
$\lambda$ in Eq.~(\ref{Psi}) as a variational parameter. This ensures a
smooth transition from the CTG ($\lambda = 1$) to the Wigner crystal
($\lambda \gg 1$), provides a strict upper bound on $\varepsilon(n)$,
and can be done semi-analytically. Indeed, the energy density of the
state $\Psi(\lambda)$ is the sum of the kinetic
$\varepsilon_{kin}^{var}$ and the potential $\varepsilon_{int}^{var}$
terms. By virtue of a formula similar to Eq.~(\ref{xc_energy}),
$\varepsilon_{kin}^{var}$ can be computed differentiating the known
energy density $\varepsilon_{\rm CS}(\lambda)$ in the
Calogero-Sutherland model~\cite{Sutherland_71} with respect to its
coupling constant $c = \lambda (\lambda - 1)$,
\begin{equation}
\varepsilon_{kin}^{var} = \left(1 - c \frac{\partial}{\partial c}\right)
\varepsilon_{\rm CS} =
\frac{\pi^2}{6} \frac{\hbar^2 n^3}{m} \frac{\lambda^3}{2 \lambda - 1},
\label{energy_kin_var}
\end{equation}
To get $\varepsilon_{int}^{var}$, we calculate it at $\lambda = \frac12,
1, 2$, and $\infty$ using the exact cluster functions
$h(x)$~\cite{Mehta_book} and interpolate between the obtained four
values by a cubic polynomial in $\lambda^{-1}$,
%
$\varepsilon_{int}^{var} = (e^2 / \kappa) n^2 [\ln (R n) - A_W -
{a_1}{\lambda}^{-1} - {a_2}{\lambda}^{-2} - {a_3}{\lambda}^{-3}]$.
%
%
For example, in the ${\cal L} \to \infty$ limit we find $a_1 \approx
-0.3173$, $a_2 \approx -0.02363$, and $a_3 \approx 0.003048$. Smallness
of $a_2$ and $a_3$ implies a high numerical accuracy of this polynomial
fit. Minimizing $\varepsilon_{kin}^{var} + \varepsilon_{int}^{var}$ with
respect to $\lambda$ (numerically), we get $\varepsilon(n)$. The quality
of our variational method can be judged by how well it compares with
Eq.~(\ref{energy_Wigner}) in the $r_s \gg 1$ limit. It is easy to see
that the functional form of $\varepsilon(n)$ is reproduced correctly,
but the coefficient in front of the phonon term is approximately
$1.022$, i.e. higher than $C_{ph}$ by mere 0.4\%. The results of this
procedure, implemented for several values of $a_B / R$, are plotted in
Fig.~\ref{Fig_energy} (b) and (c). The curves produced by the Bethe
ansatz and the variational method match virtually seamlessly. Thus, the
proposed scheme gives a theoretically well-founded and numerically
accurate DFT needed in applications, some of which are discussed next.

\noindent{\it Implications.\/}--- The main physical omission of our
theoretical model is the screening of Coulomb interactions by other 1D
subbands that may be present in a wire. Such a screening is averted if
$\kappa$ exceeds a certain threshold $\kappa_{th}$. For CN, we
estimate $\kappa_{th} \sim N {\cal L}$, e.g., $\kappa_{th} \sim 12$ for
$N = 4$ and ${\cal L} = 3$. Note that $\kappa$ is equal to the
dielectric constant $\kappa_0$ of the medium if the nanotube is immersed
in it and is equal to $(\kappa_0 + 1) / 2$ if the medium is used as a
substrate. If $\kappa < \kappa_{th}$, our theory can still apply
at sufficiently low $n$, e.g., in the Wigner crystal regime.

One possible application of our results for $\varepsilon(n)$ is a
fine-tuning of the operational parameters of carbon nanoelectronic
devices~\cite{Tans_98,Javey_02}. On a crude level, such devices are tiny
capacitors made of CN and control metallic gates. Precise knowledge of
their capacitance per unit length $C$ is desirable for their optimal
design and efficiency. The quantum and many-body effects influence the
measured value of $C$ according to the equation (see,
e.g.,~Ref.\onlinecite{Pomorski_04})
\begin{equation}
   C^{-1} = C_0^{-1} + ({\kappa}/{e^2}) \chi^{-1},\quad
  \chi^{-1} = {\partial^2 \varepsilon}/{\partial n^2},
\label{C}
\end{equation}
where $C_0^{-1} \sim (2 / \kappa) \ln (2 D / R)$ is the inverse classical
(geometric) capacitance and $\chi^{-1}$ is the inverse thermodynamic
density of states (ITDOS). The quantum correction due to the ITDOS may
be nonnegligible if the distance $D$ between between the CN and the gate
is small or if $n$ is low, so that $D \sim a$. The measurable signature
of a finite $\chi^{-1}$ would be the $n$-dependence of $C$. Recently,
capacitance of CN and their junctions was studied in
Ref.~\onlinecite{Pomorski_04} by a 3D DFT. It would be interesting to apply
our theory to the same structures for comparison.

The sign of the ITDOS is determined by the convexity of $\varepsilon(n)$
curve. From Fig.~\ref{Fig_energy} we see that at low enough electron
densities ITDOS becomes negative. This phenomenon is a generic feature
of a strongly correlated electron matter~\cite{Dolgov_81}. Unlike the
case of neutral systems, here the negative ITDOS does not imply any
thermodynamic instability but leads instead to a small {\it
overscreening\/} of an external electric charge. One possible technique
to detect such an overscreening experimentally is the scanned probe
imaging of the electrostatic potential along an ultrathin wire (e.g.,
the CN~\cite{Scanning}) set on a dielectric substrate. Above the puddles
of the electron liquid induced by stray random charges, one would see
the potential of a ``wrong'' curvature: higher near the center of the
puddle, lower near its ends. The puddles can be intentionally created by
additional small gates.

Finally, from $\varepsilon(n)$ one can extract the $n$-dependence of the
LL parameters that influence charge tunneling and low-temperature
transport in 1D wires. Preliminary results and their comparison with
other work in the literature~\cite{Hausler_02} have been reported in
Ref.~\onlinecite{Fogler_04}. A more detailed investigation that
incorporates the results derived in this paper will be presented
elsewhere~\cite{Fogler_unpublished}.

\noindent{\it Acknowledgemets.\/}--- This work was initiated at MIT and
completed at UCSD. The support from MIT Pappalardo Program, Sloan
Foundation, and C. \& W. Hellman Fund is gratefully acknowledged. I
thank L.~Levitov for important contributions to this project and also
D.~Arovas and E.~Pivovarov for discussions.

\end{document}